\documentclass[12pt]{article}

\usepackage{graphicx}

\begin{document}

\def\D3{\overline{\rm D3}}

\begin{titlepage}
\setcounter{page}{1} \baselineskip=15.5pt \thispagestyle{empty}
\begin{flushright}
\parbox[t]{2in}{
COLO-HEP-537}
\end{flushright}

\vfil

%\vspace*{0.5in}

\begin{center}
{\LARGE Nonsupersymmetric brane vacua in stabilized compactifications}
\end{center}
\bigskip

\begin{center}
%
%{\large
{Charles Max Brown and Oliver DeWolfe}
\end{center}
\begin{center}
\textit{Department of Physics, 390 UCB,
     University of Colorado,
     Boulder, CO 80309, USA}
\end{center}
\bigskip \bigskip \bigskip \bigskip

\begin{center}
{\bf
Abstract} \end{center}

\noindent We derive the equations for the nonsupersymmetric vacua of D3-branes in the presence of nonperturbative moduli stabilization in type IIB flux compactifications, and solve and analyze them in the case of two particular 7-brane embeddings at the bottom of the warped deformed conifold.  In the limit of large volume and long throat, we obtain vacua by imposing a constraint on the 7-brane embedding.  These vacua fill out continuous spaces of higher dimension than the corresponding supersymmetric vacua, and have negative effective cosmological constant.  Perturbative stability of these vacua is possible but not generic.  Finally, we argue that $\D3$-branes at the tip of the conifold share the same vacua as D3-branes.

\vfil
\begin{flushleft}
\today
\end{flushleft}

\end{titlepage}

\section{Introduction}

Compactifications with fluxes and branes provide an opportunity to construct four-dimensional vacua of string theory with many phenomenologically necessary features, such as reduced supersymmetry, hierarchies of scales and potentially a positive cosmological constant.  To make these models viable, however, one must stabilize any moduli fields that remain in the effective theory.  The most well-studied scenario takes place in type IIB string theory, where three-form fluxes can lift the complex structure moduli and the dilaton \cite{GKP, OtherFlux}.  The K\"ahler moduli, however, remain massless unless additional effects occur.  Due to the freedom of the overall volume, these effective theories are called ``no-scale".
In addition, D3-branes filling noncompact spacetime and sitting at points on the compact space feel no potential.  D3-branes are of great interest in flux compactifications, as they can provide gauge groups for a braneworld scenario or provoke cosmological evolution via brane inflation.  A lack of a brane potential in the no-scale models would have substantial consequences for any low-energy model.

The most prominent mechanism for stabilizing the K\"ahler moduli is non-perturbative strong-coupling dynamics taking place on 7-branes (or Euclidean D3-branes) wrapping 4-cycles, as described by KKLT \cite{KKLT}.  With a mild fine-tune of parameters, the geometry can be stabilized at large volume, where subleading corrections can be neglected.  In addition, the nonperturbative physics is sensitive to the locations of D3-branes, and consequently they acquire a potential as well.  It is this potential that we consider in this paper.

The general contribution of the D3-branes to the nonperturbative superpotential was formulated in \cite{Baumann}; see also \cite{More}.  In \cite{DMSU}, the general equations constraining supersymmetric vacua were obtained, and then studied in the particular case of the tip of the Klebanov-Strassler warped throat \cite{KS}, for a number of different 7-brane embeddings.  It was found that depending on the embedding, one could find for the D3-branes a moduli space of supersymmetric vacua, isolated supersymmetric vacua, or no supersymmetric vacua at all. 

In this paper, we continue to investigate the vacua of moduli-stabilized D3-branes by turning to the nonsupersymmetric case.   We consider the general equations for a minimum of the coupled system of a D3-brane and a single K\"ahler modulus.  We describe the stabilization of the K\"ahler axion in all generality, and demonstrate how large-volume solutions have compact volumes and cosmological constants approaching the results for the supersymmetric cases.

 We then turn to studying the vacua in the particular cases of two 7-brane embeddings, the so-called ``simplest Kuperstein" \cite{Kuperstein} and ``Karch-Katz" \cite{KarchKatz} embeddings.  We find that nonsupersymmetric vacua may exist at the tip of the KS throat, but they are not generic; in general one parameter of the 7-brane embedding must be tuned.  When they do exist, we study them in the most trustworthy limit of large volume and a long throat, and find that as the 7-brane embedding is varied, continuous spaces of nonsupersymmetric anti-de Sitter vacua interpolate between the supersymmetric vacua.  In our examples the supersymmetric vacua lie at fixed or partially fixed loci of the unbroken geometric symmetry, while the nonsupersymmetric vacua do not; hence the nonsupersymmetric examples, when they exist, end up filling out higher-dimensional spaces of solutions than the supersymmetric ones.
 
We consider also the issue of the stability of these nonsupersymmetric vacua; since they are anti-de Sitter, this is determined by the Breitenlohner-Freedman bound.  We find that although stability is by no means generic, there are 7-brane embeddings that produce stable nonsupersymmetric vacua.  The nonsupersymmetric vacua have cosmological constants slightly more negative than the stable supersymmetric cases, a situation not unusual in AdS supergravities.  We also extend the argument from \cite{DMSU}, proven there in the supersymmetric case, that $\D3$-branes will have the same vacua as D3-branes at the bottom of a KS throat, removing a potential obstacle to brane inflation (for some work in the context of these nonperturbative potentials see \cite{Inflation}).

In section \ref{ReviewSec} we review the moduli-stabilizing KKLT superpotential and the geometry of the conifold.  In section \ref{GeneralSec} we review the supersymmetric solutions and describe the equations for nonsupersymmetric vacua in general, before specializing to two particular embeddings and their solutions in sections \ref{KupSec}
and \ref{KKSec}.  Finally we argue that the $\D3$-branes share the same vacua as D3-branes at the tip of the conifold throat in section \ref{AntibraneSec}, before concluding in section \ref{ConclusionsSec}.

\section{Review}
\label{ReviewSec} 

\subsection{Moduli and superpotential}

We will be concerned with a flux compactification of type IIB string theory with a single complex K\"ahler modulus $\rho$ and three complex moduli for the position of a D3- (or $\D3$-) brane; the complex structure moduli and dilaton are assumed already stabilized by three-form fluxes.  The real and imaginary parts of the K\"ahler modulus are
\begin{eqnarray}
\rho = {1 \over 2} ( e^{4u} + \gamma k(Y,\overline{Y})/3) + i b \,,
\end{eqnarray}
where $b$ is the axion field associated to $C_4$, $e^{4u}$ parameterizes the volume of the corresponding 4-cycle, and $k(Y,\overline{Y})$ is the geometric (``little") K\"ahler potential for the Calabi-Yau space depending on the holomorphic coordinates $Y^I$, $I = 1,2, 3$ and their conjugates, and  $\gamma = T_{D3} \kappa_4^2$.  The associated total K\"ahler potential for all the relevant moduli is
\begin{equation}
\label{KahlerPot}
K=-3\log e^{4u}=-3\log(\rho+\overline{\rho}- \gamma k(Y,\overline{Y})/3)\,,
\end{equation}
and in what follows we will absorb $\gamma$ into $k$ for simplicity of notation.
We take as our superpotential the KKLT form with a nonperturbative contribution \cite{KKLT, Baumann},
\begin{equation}
W=  W_0+A_0e^{-a\rho}f(Y)^{1/n} \,,
\label{SuperPotential}
\end{equation}
where $W_0$ and $A_0$ are complex constants and $f(Y)=0$ defines the embedding of the $n$ 7-branes (or Euclidean D3-branes) producing the nonperturbative effects.  It is also convienent to define
\begin{equation}
A(Y) \equiv A_0 f(Y)^{1/n} \,, \quad \quad \zeta(Y)\equiv-\frac{1}{n}\log f(Y) \,,
\end{equation}
such that
\begin{eqnarray}
W=  W_0+A(Y) \, e^{-a\rho} = W_0 + A_0 e^{-a \rho - \zeta(Y)} \,.
\end{eqnarray}
In addition, we will find it useful to introduce the (in general complex) quantity
\begin{eqnarray}
\omega(\rho, Y) \equiv {W_0 \over A(Y)} e^{a \rho} \,,
\end{eqnarray}
which measures the relative magnitude of the perturbative and nonperturbative terms in the superpotential.  We note that the quantities $k(Y, \bar{Y})$, $\rho$ and $\zeta(Y)$ are not uniquely defined, but transform according to ``little K\"ahler
transformations'',
\begin{eqnarray} \nonumber
k &\to& k + 3~ \xi(Y) + 3~ \bar{\xi}(\bar{Y}) \,, \\
\label{LittleKahler}
\rho &\to& \rho + \xi(Y) \,, \\
\zeta &\to& \zeta - a \, \xi(Y) \,,
\nonumber
\end{eqnarray}
where $e^{4u}$ and $a \rho + \zeta(Y)$ are invariants.

\subsection{The conifold and its tip}

We will study the geometry of the deformed conifold, and in particular its tip.
The deformed conifold may be defined by a set of four complex variables $z^1$, $z^2$, $z^3$, $z^4$ with the constraint
\begin{eqnarray}
\sum_{A=1}^4 (z^A)^2 = \epsilon^2 \,,
\end{eqnarray}
where $\epsilon$ determines the degree of deformation; for convenience, we will choose it to be real and positive.  A geometric $SO(4)$ symmetry acts on the $z^A$ in the obvious way.  We can think of the conifold as a compact five-dimensional space times a radial coordinate,
\begin{equation}
r^3 \equiv  \sum^{4}_{A=1} \, |z^A|^2 \,,
\end{equation}
where at $r \to \infty$ the space approaches the ordinary conifold with metric
\begin{eqnarray}
ds^2 = dr^2 + r^2 d \Omega^2_{T^{1,1}} \,,
\end{eqnarray}
showing explicitly the space is a cone over the compact five-dimensional space $T^{1,1}$, which has topology $S^2 \times S^3$.  Meanwhile we can see that the ``bottom" or ``tip" of the throat occurs at the minimal value
\begin{eqnarray}
r^3= \epsilon^2 \,, \quad \quad @ \ {\rm tip} \,,
\end{eqnarray}
at which point the $S^2$ shrinks to zero size but the $S^3$ remains; the $SO(4)$ symmetry acts naturally on this $S^3$.  In terms of the radial variable $\tau$ defined by $r^3 = \epsilon^2 \cosh \tau$, the metric near the tip becomes
\begin{eqnarray}
ds^2 \approx d\tau^2 + \tau^2 d \Omega^2_{S^2}  + d \Omega_{S^3}^2 \,.
\end{eqnarray}
The tip of the throat corresponds to taking the $z^A$ to be real,
\begin{eqnarray}
z^A = |z^A| \,, \quad \quad @ \ {\rm tip} \,.
\end{eqnarray}
In calculations it is often necessary to choose three of the four $z^A$ as independent variables.  We will take $z^1$ to be dependent, so that
\begin{eqnarray}
z^1 = \sqrt{\epsilon^2 - (z^2)^2 - (z^3)^2- (z^4)^2} \,, \quad \quad {\partial z^1 \over \partial z^a} = - {z^a\over z^1} \,, \quad a = 2,3, 4\,,
\end{eqnarray}
which is valid as long as $z^1 \neq 0$.  Near the tip the little K\"{a}hler potential has  the form \cite{Comments},
\begin{equation}
k(Y,\overline{Y})\approx k_0+ Q \left[ \left({r^3\over \epsilon^2} - 1\right) - {1 \over 10 }  \left({r^3 \over \epsilon^2}-1\right)^2  + \ldots \right] \,,
\label{LittleKahler}
\end{equation}
where $k_0$ and $Q$ are constants, the latter taking the form
\begin{eqnarray}
Q \equiv  {2^{1/6} \over 3^{1/3}} T_{D3} \kappa_4^2 \epsilon^{4/3} \,.
\end{eqnarray}
The constant $k_0$ is absent from $K$ (equation (\ref{KahlerPot})) and its derivatives, which only depend on the little K\"ahler-invariant $e^{4u}$, and thus appears only in the superpotential in the combination $A_0 e^{-a k_0/6}$.  For convenience, we will absorb it into our definition of $A_0$, and so set $k_0 = 0$ in the following.

\section{General Solutions}
\label{GeneralSec}

In this section we first review the equations and solutions for supersymmetric solutions for the K\"ahler and brane moduli, and then develop the equations for nonsupersymmetric vacua in generality.  In the following sections, we pick explicit choices for the 7-brane embedding and study the corresponding solutions.

\subsection{Supersymmetric Solutions}

The potential for the moduli in the presence of the superpotential is as usual
\begin{equation}
V=e^K \left(g^{\alpha\overline{\beta}}D_\alpha WD_{\overline{\beta}}\overline{W}-3 |W|^2 \right) \,,
\end{equation}
where $\alpha, \beta = \rho, I$ runs over the four moduli, with $I$ taking the three values for the geometric moduli, and as usual the K\"ahler covariant derivative is
\begin{eqnarray}
D_\alpha W = \partial_\alpha W + W \partial_\alpha K \,.
\end{eqnarray}
For the case at hand given by equations (\ref{KahlerPot}), (\ref{SuperPotential}), these evaluate at the tip of the geometry to
\begin{eqnarray}
D_\rho W &=& - {W_0 \over \omega} ( a + 3 e^{-4u} (1 + \omega)) \,, \\
D_I W &=& {W_0 \over \omega} \partial_I \zeta  \,.
\end{eqnarray}
The supersymmetric solutions satisfy $D_\rho W = D_I W = 0$, and are thus given by \cite{DMSU} 
\begin{eqnarray}
 \omega =-1-{a  e^{4u} \over 3}\,, \quad \quad
\label{SusySolution}   \partial_I \zeta(Y)=0 \,.
\end{eqnarray}
The second equation is equivalent to $(\partial_I f)/f = 0$, which in principle fixes the geometric moduli $Y^I$, although in practice many cases have moduli spaces of solutions, as we shall discuss further.   Meanwhile the right-hand-side of the first equation is real, so given specified values of the $Y^I$ the axion $b \equiv {\rm Im}\ \rho$ is fixed by the imaginary part of this relation as
\begin{equation} 
b=-\frac{1}{a}\arg \left(-\frac{W_0}{A(Y)} \right) \,,
\label{SusyImRho}
\end{equation}
while the volume $e^{4u}$ is stabilized by the real part of the equation at
\begin{eqnarray}
\left|W_0 \over A(Y) \right| e^{a e^{4u} /2} = 1 + {a e^{4u} \over 3} \,. 
\end{eqnarray}
As is well-known, large volume solutions can only exist for sufficiently small $|W_0/A(Y)|$.

\subsection{Nonsupersymmetric solutions}

In order to find non-supersymmetric vacua we will identify all extrema $\partial_{\alpha}V=0$, ruling out the cases which satisfy equation (\ref{SusySolution}) and are therefore supersymmetric. It is useful to calculate the metric on moduli space $g_{\alpha \bar\beta} \equiv \partial_\alpha \partial_{\bar\beta} K$ and its inverse $g^{\alpha \bar\beta}$, which at the tip of the geometry where $\partial_I k = 0$ take the form,
\begin{eqnarray}
\label{metric}
g_{\alpha \bar\beta} = \left(
\begin{array}{cc}
	3 e^{-8u} & 0  \\
	0 & e^{-4u} k_{I \bar{J}}
\end{array}
\right) \,, \quad \quad
g^{\alpha \bar\beta} = \left(
\begin{array}{cc}
	\frac{1}{3} e^{8u} & 0  \\
	0 & e^{4u} k^{I \bar{J}}
\end{array}
\right) \,,
\end{eqnarray}
where $k_{I\bar{J}} \equiv \partial_I \partial_{\bar{J}} k$ and $k^{I \bar{J}}$ is its inverse.  For the particular case of the $z^a$ coordinates at the tip these become
\begin{eqnarray}
k_{a\bar{b}} = {Q\over \epsilon^2} \left( \delta_{a \bar{b}} + {z^a z^{\bar{b}} \over |z_1|^2} \right) \,, \quad \quad
k^{a\bar{b}} = {\epsilon^2 \over Q} \left( \delta^{a \bar{b}} - {z^a z^{\bar{b}} \over \epsilon^2}\right) \,.
\end{eqnarray}
The value of the potential at the tip is then
\begin{eqnarray}
V = {a|W_0|^2 e^{-8u} \over |\omega|^2} \left( {a e^{4u} \over 3} + 2 + \omega + \bar\omega + G(Y) \right) \,,
\end{eqnarray}
where we have defined the {\em real} quantity,
\begin{eqnarray}
G(Y) \equiv {1 \over a} k^{I \bar{J}} \partial_I \zeta \partial_{\bar{J}} \bar\zeta \,,
\end{eqnarray}
  We shall find that equations simplify if we measure $\omega$ in terms of its separation from the SUSY solution,
\begin{equation}
\delta\equiv\omega-\omega_{SUSY}=\omega+1+{a e^{4u} \over3} \,,
\end{equation}
in terms of which we have
\begin{eqnarray}
\label{VEqn}
V = {a|W_0|^2 e^{-8u} \over |\omega|^2} \left( -{a e^{4u} \over 3}  + \delta + \bar\delta + G(Y) \right) \,.
\end{eqnarray}
The supersymmetric solution then has
\begin{eqnarray}
\delta_{SUSY} = G_{SUSY} = 0 \,, \quad \quad V_{SUSY} = -{ a^2 |W_0|^2 e^{-4u} \over 3 |\omega|^2} \,.
\end{eqnarray}
The cosmological constant is manifestly negative.

We now turn to the first derivatives of the potential, the vanishing of which will give us the general vacua. For the $\rho$-derivative we find
\begin{eqnarray}
\partial_\rho V = - {a |W_0|^2 e^{-12 u} \over |\omega|^2} \left( a e^{4u} \bar\delta + 2 (\delta + \bar\delta) + (a e^{4u} +2) G \right) \,.
\end{eqnarray}
We note immediately that the only term that is not real is the first one with $\bar\delta$.  Thus upon imposing $\partial_\rho V = 0$, the imaginary part of the equation simply requires
\begin{eqnarray}
{\rm Im}\ \omega = 0 \,,
\end{eqnarray}
equivalent to the statement that the perturbative and nonperturbative terms in the superpotential have the same phase up to sign.  This implies that just as in the supersymmetric case, the axion $b$ (which appears exclusively inside $\omega$ in all our expressions) adjusts itself to ensure $\omega$ is real, resulting again in the equation (\ref{SusyImRho}).

The remaining real part of $\partial_\rho V = 0$ then requires
\begin{eqnarray}
\label{RhoEqn}
\delta(e^{4u}, Y) = - {a e^{4u} +2 \over a e^{4u} + 4} \, G(Y) \,.
\end{eqnarray}
In the supersymmetric cases, this equation is trivially satisfied by $\delta = G = 0$; the nonsupersymmetric cases will involve both sides being nonzero.

The form of the equation $\partial_a V = 0$ is more strongly dependent on the choice of the 7-brane embedding $f$.  It simplifies in the basis of the $z^a$, where it can be written as
\begin{eqnarray}
{|W_0|^2 a\over 3\omega^2 e^{8u}}  \left((2+3\delta)\partial_a \zeta + (1+{3 z^b \partial_b \zeta\over a Q}) \delta_a^{\bar{a}} \partial_{\bar{a}}\bar{\zeta}-{3k^{b\bar{c}}\partial_{\bar{c}}\bar{\zeta}\over a}(\partial_a\partial_b \zeta-\partial_a \zeta \partial_b \zeta)\right) = 0\,.
\label{aDerivEqns}
\end{eqnarray}
In general we thus have three complex equations (\ref{aDerivEqns}) and one real equation (\ref{RhoEqn}) for the geometric moduli coupled to the volume.  To proceed further, we need to pick particular forms of the 7-brane embedding.

Before turning to specific types of 7-brane embeddings, we note that for any solution valid at large volume $e^{4u} \gg 1$, equation (\ref{RhoEqn}) gives
\begin{eqnarray}
\label{DeltaG}
\delta \approx - G \,.
\end{eqnarray}
$G$ is in turn a function of the coordinates $Y$ and not directly the volume.  In the case $e^{4u} \gg G$ where the volume is much larger than this function, the expression for the potential is dominated by the volume factor, and
\begin{eqnarray}
\label{VVSusy}
V \approx V_{SUSY}  < 0\,.
\end{eqnarray}
Thus for large-volume situations, the cosmological constant for a nonsupersymmetric vacuum will be generically negative and close to the supersymmetric value.  In our examples, this will indeed be the case.

Furthermore, when $e^{4u} \gg G$ we can only satisfy (\ref{DeltaG}) if
\begin{eqnarray}
\label{SusyVol}
 \omega \approx- {a e^{4u} \over 3} \quad \to \quad \left|W_0 \over A(Y) \right| e^{a e^{4u} /2}  \approx {a e^{4u}\over 3} \,,
\end{eqnarray}
which is again the supersymmetric limit.  Thus large volume cases generically result in both the cosmological constant and the compact volume having values close to the supersymmetric values, and as with those cases, large volume can only be achieved by tuning $W_0$ sufficiently small to satisfy (\ref{SusyVol}).   Once this is done, however, the large volume solution exists and is essentially the same as the supersymmetric volume, the difference being an order-one difference between $A(Y)$ in each case.

Having discussed the nonsupersymmetric solutions in general, we turn now to two specific examples of embeddings.
 
\section{Simplest Kuperstein embedding}
\label{KupSec}

We take as our first example of a 7-brane configuration the simplest case of Kuperstein embeddings \cite{Kuperstein},
\begin{equation}
f_K=z^1-\mu \,,
\end{equation}
where $\mu$ is a parameter, in general complex; its modulus gives the minimum value of $r$ that the 7-brane reaches,
\begin{eqnarray}
r_{min} = |\mu|^{2/3} \,.
\end{eqnarray}
The $SO(4)$ symmetry of the conifold is broken to the $SO(3)$ acting on $z^2$, $z^3$ and $z^4$.
Supersymmetric vacua at the tip of the throat for this embedding were found \cite{DMSU} to live at two opposite poles of the $S^3$, $z^2 = z^3 = z^4=0$, $z^1 = \pm \epsilon$; this is an example of an embedding without a supersymmetric moduli space.  The poles are in fact the only fixed points of the surviving $SO(3)$ symmetry on the tip, and we shall find that the nonsupersymmetric vacua, when they exist, exist away from the poles, and thus in $S^2$ moduli spaces.

In what follows we shall restrict ourselves to the tip of the geometry.  We have for this embedding
\begin{eqnarray}
\partial_a \zeta = {z^a \over n f z^1} = {z^a \over n z^1 (z^1 - \mu)} \,,
\end{eqnarray}
and
\begin{eqnarray}
G = {\epsilon^2 \over a Q} \left( \delta^{a\bar{b}} - {z^a z^{\bar{b}} \over \epsilon^2} \right) {\partial_a \zeta \partial_{\bar{b}} {\bar{\zeta}}} = {\epsilon^2 - |z^1|^2 \over a Q n^2 |f|^2}  \equiv {\epsilon^2 - |z^1|^2 \over \Gamma |f|^2} \,,
\end{eqnarray}
where we used that the $z^a$ are real at the tip,  as well as $|z^1|^2 + \sum_a |z^a|^2 = \epsilon^2$; the last equality defines the constant $\Gamma$.  The $\partial_\rho V = 0$ equation  (\ref{RhoEqn}) thus gives
\begin{eqnarray}
\delta = - {a e^{4u} +2 \over a e^{4u} + 4} \, {\epsilon^2 - z^2 \over \Gamma |f|^2} \,,
\label{RhoEqn2}
\end{eqnarray}
where we have set $z \equiv z^1$. Meanwhile, the $\partial_a V = 0$ equations give us
\begin{eqnarray}
{ |W_0|^2 z^a a \over \omega^2 e^{8u} n f z^1} \left[-\delta  - {2 \over 3} - {1 \over 3} {f \over \overline{f}} + {1 \over |f|^2 \Gamma} ( f n z^1 + (n-1) (\epsilon^2 - |z^1|^2)) \right] =0\,. 
\label{KupRhoEqn}
\end{eqnarray}
We note that in the Kuperstein case the equations for supersymmetric vacua become
\begin{eqnarray}
\delta =0 \,, \quad \quad z^a = 0 \,, \quad a = 2, 3, 4 \,,
\end{eqnarray}
and hence equation (\ref{KupRhoEqn}) is trivially satisfied by $z^a = 0$.    Assuming therefore that $z^a \neq 0$ for at least one $z^a$ leads to the nonsupersymmetric cases, and we can cancel the overall factors to obtain
\begin{eqnarray}
\delta = - {2 \over 3} - {1 \over 3} {f \over \bar{f}} + {1 \over |f|^2 \Gamma} ( f n z + (n-1) (\epsilon^2 - z^2))  \,. 
\label{aEqn}
\end{eqnarray}
Note that the $z^a$ only appear in these equations in terms of $z$ as $(z^2)^2 + (z^3)^2 + (z^4)^2 = \epsilon^2 - z^2$.  $z = \pm \epsilon$ only occurs for $z^a = 0$, which is the supersymmetric case; for a general solution $0 \leq z < \epsilon$ we will have an $SO(3)$ remnant of the $SO(4)$ of the conifold acting on the $z^a$, and thus the general solution space will be an $S^2$.

We can combine (\ref{RhoEqn2}) and (\ref{aEqn}) to eliminate $\delta$, obtaining
a quadratic equation for $z$,
\begin{eqnarray}
z^2 (- \Gamma + {2 \over \tau}) + z ( {4 \over 3} \Gamma \mu + {2 \over 3} \Gamma \bar{\mu} - n \mu ) - {2 \over 3} \Gamma |\mu|^2- {1 \over 3} \Gamma \mu^2 + (n-{2 \over \tau})\epsilon^2 =0  \,,
\label{BothEqn}
\end{eqnarray}
where we have defined
\begin{equation}
\tau=ae^{4u}+4\,,
\end{equation}
which eliminates exponential dependence on $e^{4u}$; we can use (\ref{RhoEqn2}) and (\ref{BothEqn}) as our independent equations.
Let us now count degrees of freedom.  We have one real equation (\ref{RhoEqn2}) and one complex equation (\ref{BothEqn}), and two real degrees of freedom $e^{4u}$ and $z$; hence in general our system is overconstrained.  This should not be surprising since we restricted to looking at the bottom of the throat, essentially requiring $z$ to be real.  

Do solutions exist?  In general they will not, but we have a tunable parameter: $\mu$, characterizing the 7-brane embedding.  Hence we expect that for certain values of $\mu$, solutions may exist.  This turns out to be the case.
The imaginary part of (\ref{BothEqn}) reduces to a linear equation for $z$,
\begin{eqnarray}
{\rm Im}\, \mu \, \Big( z( 2 \Gamma-3n) - 2 \Gamma\, {\rm Re}\, \mu \Big)=0\,.
\label{ImEqn}
\end{eqnarray}
If we view this relation as a constraint on $\mu$, the remaining equations --- (\ref{RhoEqn2}) and the real part of (\ref{BothEqn}) --- are then solvable for $e^{4u}$ and $z$.

There are two possible constraints on $\mu$ that can be imposed.  We consider them in turn.

\subsection{Real  7-brane embedding}

The simplest constraint on $\mu$ solving (\ref{ImEqn}) is just, 
\begin{eqnarray}
{\rm Im} \, \mu = 0 \,,
\end{eqnarray}
which at the tip is equivalent to $f = \bar{f}$.  In this case (\ref{BothEqn}) reduces to
\begin{eqnarray}
z^2 (- \Gamma + {2 \over \tau}) + z ( 2 \Gamma \mu  - n \mu ) -  \Gamma \mu^2+ \epsilon^2 (n - {2 \over \tau}) =0  \,,
\label{RealPartEqn2}
\end{eqnarray}
giving the solution
\begin{eqnarray}
\label{FirstzSoln}
z=\frac{\tau \mu(2\Gamma-n) \pm \sqrt{4\epsilon^2(-2+n\tau)(-2+\Gamma\tau)+\mu^2\tau(8\Gamma+n^2\tau-4n\Gamma\tau)}}{2(\Gamma \tau-2)} \,,
\end{eqnarray}
while (\ref{RhoEqn2}) gives
\begin{eqnarray}
\delta=\frac{z^2(1-\Gamma)+\epsilon^2(n-1)+z\mu(2\Gamma-n)-\mu^2\Gamma}{\Gamma(z-\mu)^2} \,.
\end{eqnarray}
In general this is a transcendental pair of coupled equations, since $z$ depends on $\tau \equiv a e^{4u} + 4$, while $\delta$, which contains a term exponential in $e^{4u}$, depends on $z$. 

It is useful to consider the solution in  the large-volume limit $e^{4u} \to \infty$, where the expressions simplify; we shall see momentarily that this is a consistent limit as long as the parameters are chosen appropriately.  Of course, this is also the limit where the equations do not receive stringy corrections and so can be trusted, and so is the regime of most interest.  
The equation for $\delta$, unpacked in terms of $\omega$, gives for the volume
\begin{eqnarray}
\left|W_0 \over A(Y) \right| e^{a e^{4u} /2}  = 1 +{a e^{4u}\over 3} + {a e^{4u} +2 \over a e^{4u} + 4} \, {\epsilon^2 - z^2 \over \Gamma |f|^2} \quad \to \quad \left|W_0 \over A(Y) \right| e^{a e^{4u} /2}  \approx {a e^{4u}\over 3}  \,, 
\end{eqnarray}
which approaches the form of the supersymmetric expression for the volume, as argued in general in (\ref{SusyVol}).
As with the supersymmetric vacua, we can only satisfy this transendental equation in $e^{4u}$ at large volume by fine-tuning $W_0$ to be small.

Meanwhile, the solution (\ref{FirstzSoln}) for $z$ becomes
\begin{eqnarray}
\label{zSolns}
z  ={1 \over 2 \Gamma}\left( \mu (2 \Gamma - n) \pm \sqrt{ \mu^2 n^2 + 4 n \Gamma(\epsilon^2  - \mu^2)}\right) \,.
\end{eqnarray}
We require $0 \leq |z/\epsilon| < 1$ to restrict the variable to the tip; recall that $|z| = |\epsilon|$ will imply the supersymmetric solutions.   In general this will place a constraint on the magnitude of $\mu$, as certain values of $\mu$ will lead to unphysical $z$.

Let us explore this in a further limit, that of a long throat.  The parameter
\begin{eqnarray}
\Gamma \equiv a n^2 Q =2 \pi {2^{1/6} \over 3^{1/3}}  n  T_{D3} \kappa_4^2 \epsilon^{4/3} \equiv\Gamma' \epsilon^{4/3} \,, 
\end{eqnarray}
scales with a power of $\epsilon$ and hence captures the depth of the throat; the last relation above extracts the power of $\epsilon$ and defines the pure number $\Gamma'$.  The regime of $\epsilon$ small is the limit under best control, with a long throat having a tip far from the rest of the geometry.  In this limit $\Gamma \to 0$; the solution to (\ref{zSolns}) with the minus sign goes to infinity as $1/\Gamma$ and hence is unphysical, while the solution with the plus sign asymptotes to\footnote{To be precise, to achieve this result we have taken an expansion in $\epsilon^{4/3} (1 -\epsilon^2/\mu^2)$, but for $\epsilon \ll 1$ this quantity is always small for the ranges of $\mu$ permitted in  (\ref{MuRange}).}
\begin{eqnarray}
\label{z+}
z  =  {\epsilon^2 \over \mu} - {\Gamma' \epsilon^{4/3} \mu \over n}\,. 
\end{eqnarray} 
The solution (\ref{z+}) only exists for $0 < |z| < \epsilon$, leading to the requirements
\begin{eqnarray}
\label{MuRange}
\epsilon < \mu <  {n\epsilon^{-1/3}  \over \Gamma'}\,.
\end{eqnarray}
As $\mu$ approaches the edges of the acceptable range, $ z \to \epsilon$ and the nonsupersymmetric vauca approach one of the supersymmetric ones; within the range, we have nonsupersymmetric vacua filling out an $S^2$ and some intermediate value of $z$.\footnote{Strictly speaking, our coordinate choice to eliminate $z = z^1$ in terms of the other variables breaks down at the one value $\mu = \epsilon^{1/3} \sqrt{n/\Gamma'}$ that gives $z=0$.}

\begin{figure}
\begin{center}
\includegraphics%[width=0.5\textwidth],clip]
{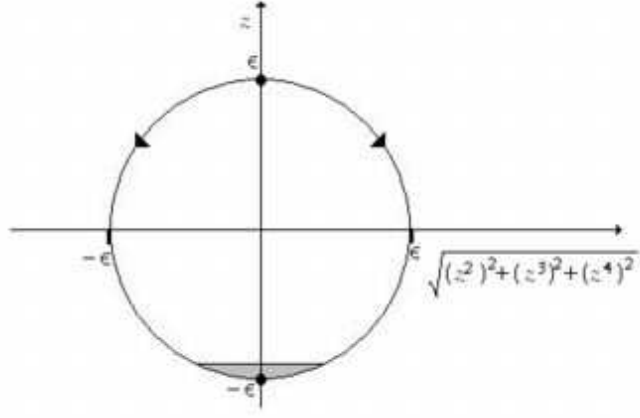}
\caption{A cross-section of the tip with supersymmetric vacua at the poles and nonsupersymmetric vacua interpolating between as $\mu$ is varied.
\label{fig:Kup}}
\end{center}
\end{figure}

Hence we see that for the regime of most theoretical control, namely large volume and a long throat, a two-sphere of nonsupersymmetric vacua appears along with the pointlike supersymmetric vacua (see figure~\ref{fig:Kup}).  

Turning now to the four-dimensional cosmological constant,  
we consider  the value of the potential (\ref{VEqn}) at these minima in the large-volume limit.   At large volume we have both $\delta$ and $G$ showing up at zeroth order in $e^{4u}$, which are subleading, so as in (\ref{VVSusy}) we obtain $V \approx V_{SUSY}$.
Including the subleading terms, using $\delta \approx  - G$, we get
\begin{eqnarray}
V \approx 
- {a |W_0|^2 e^{-8u} \over \omega^2} \left( {a e^{4u} \over 3} +  {1 \over \Gamma' } {\epsilon^2 -z^2 \over \epsilon^{4/3} (z - \mu)^2} + \ldots \right)
= V_{SUSY} - {a |W_0|^2 e^{-8u} \over \omega^2}{1 \over \Gamma' } {\epsilon^2 -z^2 \over \epsilon^{4/3} (z - \mu)^2} + \ldots
\end{eqnarray}
We thus find that the nonsupersymmetric vacua in this limit have (slightly) {\em lower} cosmlogical constant than the supersymmetric ones. 

This may seem unusual, but is actually common for supergravities with AdS vacua.  In particular, in 5D maximally supersymmetric gauged supergravity, which consists of the lowest-mass modes of the compactification of type IIB supergravity on $AdS_5 \times S^5$, the maximally supersymmetric vacuum with all scalars vanishing is actually a global {\em maximum} of the potential, while other solutions with less SUSY or no SUSY exist with more negative values of $V$, both stable and unstable.

Finally, one is curious about the stability of these vacua, which requires calculating the second derivatives; we do so here in the limits
 $e^{4u} \to \infty$ and $\epsilon \to 0$.  Taking advantage of the $S^2$ symmetry we will evaluate the second derivatives at the particular point
\begin{eqnarray}
(z^2)^2=(z^3)^2=(z^4)^2={\epsilon^2-(z)^2 \over 3} \,,
\end{eqnarray}
where $z$ is the solution given in equation (\ref{z+}).  To facilitate our series expansion of the second derivatives in powers of $\epsilon$ we will first consider the case where $\mu$ is some fraction of its maximum value
\begin{eqnarray}
\mu=\beta \mu_{max} = {\beta n \epsilon^{-1/3} \over \Gamma'} \,, \quad \quad |\beta| \leq 1 \,. 
\end{eqnarray}
In this limit we have evaluated the leading order terms of the second derivatives with respect to the holomorphic variables as
\begin{eqnarray}
\label{SKup2d}
V_{\rho \rho}= {a^3 \over 3 e^{8u}} \,,
&&V_{\rho a} =  {-a^2 \Gamma' \epsilon^{1/3} \sqrt{1-\beta^2} \over 3\sqrt{3} e^{8u} n^2 \beta^2}\,, \\
V_{a a} = {-a \Gamma' (27+\beta^2+2\beta^4)\over 135 e^{8u} n^2 \beta^4 \epsilon^{2/3}}\,, 
&&V_{a b}|_{a \neq b} = { a \Gamma' (-27 +26 \beta^2 + \beta^4) \over 135 e^{8u} n^2 \beta^4 \epsilon^{2/3}}\,, \nonumber
\end{eqnarray}
where $V_{\alpha \beta}\equiv \partial_\alpha \partial_\beta V$; in this case $V_{\bar{\alpha} \bar{\beta}} \equiv (V_{\alpha\beta})^* = V_{\alpha\beta}$.  The mixed second derivatives are a bit more complicated,
\begin{eqnarray}
&&V_{\rho \bar{\rho}}={a^4 \over 3 e^{4u}} \,, \quad \quad \quad
V_{\rho \bar{a}}={a^3 \Gamma' \epsilon^{1/3} \sqrt{1-\beta^2} \over 3\sqrt{3} e^{4u} n^2 \beta^2 }\,,
\nonumber \\
&&V_{a \bar{a}}={a^2{\Gamma'}^2\epsilon^{2/3}(1-\beta^2) \over 9 e^{4u} n^4 \beta^4}+{2 a \Gamma'(-9+23\beta^2+\beta^4) \over 135 e^{8u} n^2 \beta^4\epsilon^{2/3}}\,, \\
&&V_{a\bar{b}}|_{a \neq b}={a^2 {\Gamma'}^2 \epsilon^{2/3}(1-\beta^2) \over 9 e^{4u}n^4 \beta^4} + {a \Gamma'(-18+19\beta^2-\beta^4) \over 135 e^{8u} n^2 \beta^4 \epsilon^{2/3}}\,. \nonumber
\end{eqnarray}
Since the vacua have $V < 0$, they will be  in AdS space, and thus stability is determined by the Breitenlohner-Freedman bound, which in 4D has the form
\begin{eqnarray}
{m^2 \over |V|}\geq -{3 \over 4}\,,
\end{eqnarray}
where $m^2$ represents each of the eight eigenvalues of the mass matrix 
\begin{eqnarray}
M^2=\left(
\begin{array}{cc}
g^{\gamma \bar\beta}V_{\alpha \bar\beta} & g^{\gamma \bar{\beta}}V_{\bar\alpha \bar\beta} \\
g^{\bar\gamma \beta}V_{\alpha \beta} & g^{\bar\gamma \beta}V_{\bar\alpha \beta}
\end{array}
\right)\,.
\end{eqnarray}
We can evaluate the eigenvalues of $M^2/|V|$ to be
\begin{eqnarray}
\label{SKupM/V}
{m^2\over|V|}=\left\{{a^2 e^{8u}\over3}, {a^2 e^{8u}\over 3}, {11\over15}+{3\over5\beta^2}, {2\over15}+{6\over5\beta^2}, {2\over15}+{6\over5\beta^2}, 0, 0,3-{3\over \beta^2} \right\}\,.
\end{eqnarray} 
So if ${2/\sqrt{5}}\leq |\beta| \leq 1$ the last eigenvalue is greater than ${-3/4}$ and all of the eigenvalues satisfy the Breitenlohner-Freedman bound, and therefore we have stable non-supersymmetric vacua. 

If we expand $\mu$ around its minimum 
\begin{eqnarray}
\mu=\beta\mu_{min}=\beta\epsilon \,, \quad \quad|\beta| \geq 1 \,. 
\end{eqnarray}
We can evaluate the eigenvalues to be
\begin{eqnarray}
\nonumber
{m^2\over|V|}&=&\Bigg\{ {a^2 e^{8u}\over 3}+{6\over{\Gamma'}^2 \epsilon^{8/3}(\beta^2-1)^2}, {a^2 e^{8u}\over 3}+{6\over{\Gamma'}^2 \epsilon^{8/3}(\beta^2-1)^2}, {6 n^2 (4+\beta^2)\over 5 {\Gamma'}^2 \epsilon^{8/3} (\beta^2-1)},\\
&&  {6 n^2 (4+\beta^2)\over 5{\Gamma'}^2 \epsilon^{8/3} (\beta^2-1)}, {3(-5+n^2(4+\beta^2))\over 5{\Gamma'}^2 \epsilon^{8/3}(\beta^2-1)^2}, 0, 0,  {-3(1+n^2\beta^2) \over{\Gamma'}^2 \epsilon^{8/3}(\beta^2-1)^2} \Bigg\} \,.
\end{eqnarray}
Since the last eigenvalue is large and negative for all values of $\beta$ we see that there are not stable vacua near $\mu=\epsilon$.  Similarly, we also considered the intermediate value of $\mu$ being of order 1 and found the eigenvalues
\begin{eqnarray}
{m^2\over |V|}=\Big\{ {a^2 e^{8u}\over 3}, {a^2 e^{8u}\over 3}, {6n^2\over5\Gamma'^2\mu^2\epsilon^{2/3}},{6n^2\over5\Gamma'^2\mu^2\epsilon^{2/3}},{3n^2\over\Gamma'^2 \mu^2 \epsilon^{2/3}},0,0,{-3n^2\over\Gamma'^2 \mu^2 \epsilon^{2/3}} \Big\} \,,
\end{eqnarray}
where the last one is seen to be large and negative.  Thus in most of these regimes no stable nonsupersymmetric vacua exist, but stable vacua do appear for $\mu$ near its upper bound.

\subsection{Complex 7-brane embedding}

Consider now the other possible constraint on $\mu$; this is more complicated as it involves $z$ as well:
\begin{eqnarray}
\label{ReMuEqn}
{\rm Re}\, \mu = z \left( 1 - {3n \over 2 \Gamma} \right) \,.
\end{eqnarray}
Strictly speaking we should be thinking of ${\rm Re} \, \mu$ being fixed and this constituting an extra constraint on $z$.  However, mathematically it is convenient to eliminate ${\rm Re} \, \mu$ from the quadratic equation for $z$.  Terms linear in $z$ disappear leaving us with
\begin{eqnarray}
\label{ReMuSoln}
z^2 = {\left( \Gamma ({\rm Im}\, \mu)^2 /3 + \epsilon^2( 2 /\tau - n)\right) \over \left( - 3 n^2 / 4 \Gamma+ 2 /\tau - n\right)} \,.
\end{eqnarray}
Together (\ref{ReMuEqn}) and (\ref{ReMuSoln}) produce a constraint on $\mu$ in terms of $\Gamma$ and $n$ only:
\begin{eqnarray}
({\rm Re}\, \mu)^2 \left( {2 \over \tau} - n - {3 n^2 \over 4 \Gamma} \right) = \left( {1 \over 3} \Gamma ({\rm Im}\, \mu)^2 + \epsilon^2( {2 \over \tau} - n) \right)\left( 1 - {3 n \over \Gamma} + {9 n^2 \over 4 \Gamma^2} \right) \,,
\label{MuEqnOnly}
\end{eqnarray}
and given this constraint the solution for $z$ comes simply from (\ref{ReMuEqn}).
Let us consider the limit of large volume ($\tau \to \infty$) and long throat ($\Gamma \to 0$) without making an assumption about the magnitude of $\mu$ relative to $\Gamma$.  In this case (\ref{MuEqnOnly}) reduces to
\begin{eqnarray}
\label{MuValues}
({\rm Re}\, \mu)^2 + ({\rm Im}\, \mu)^2 \equiv |\mu|^2 = {3 n \over \Gamma'}  \epsilon^{2/3} \,.
\end{eqnarray}
Hence as with the ${\rm Im} \, \mu = 0$ case, where $|\mu|$ was bounded below at $\epsilon$ and above at $\epsilon^{-1/3}$, this solution also involves a constraint on the magnitude of $|\mu|$, in this case determining the exact value.  We find for $z$,
\begin{eqnarray}
z = - {2 \Gamma' \epsilon^{4/3} \over 3n} \, {\rm Re} \, \mu\,.
\end{eqnarray}
For a typical solution of (\ref{MuValues}) we have ${\rm Re}\, \mu \sim \epsilon^{1/3}$, meaning
\begin{eqnarray}
z \sim \epsilon^{5/3} \,,
\end{eqnarray}
and the solution is driven towards $z\sim 0$.  Thus these solutions approach the maximal $S^2$ at the tip.  In the particular case ${\rm Im} \, \mu = 0$, we find
\begin{eqnarray}
z = -2 \epsilon \sqrt{\Gamma \over 3n} = - 2 {\epsilon^2 \over \mu} \,,
\end{eqnarray}
which also coincides with a real 7-brane embedding solution  from the previous subsection with the particular value ${\rm Re}\ \mu = \sqrt{3n/\Gamma'} \epsilon^{1/3}$.

Studying the issue of stability, if we set
\begin{eqnarray}
({\rm Im} \ \mu)^2={\beta 3 n \epsilon^{2/3} \over \Gamma'}\,, \quad \quad
({\rm Re} \ \mu)^2={(1 - \beta) 3 n \epsilon^{2/3} \over \Gamma'}\,,
\end{eqnarray}
with $0 \leq \beta \leq 1$, we find the eigenvalues of the mass matrix $M^2/|V|$ to be
\begin{eqnarray}
{m^2\over |V|}= \Bigg\{{a e^{8u}\over 3},{a e^{8u}\over 3},{2n\over5 \Gamma' \epsilon^{4/3}},{2n\over5 \Gamma' \epsilon^{4/3}}, {n\over5\Gamma' \epsilon^{4/3}}, 0, 0, -{n\over5\Gamma' \epsilon^{4/3}} \Bigg\} \,,
\end{eqnarray}
independent of $\beta$.  Since the last eigenvalue is large and negative, these vacua are unstable.  

In summary, we have found that with a constraint on the embedding parameter $\mu$ we can generate $S^2$'s of nonsupersymmetric vacua in the large-volume limit, in some cases stable ones.  We turn now to a second example of a 7-brane embedding, which shares several features with the Kuperstein case.

\section{Karch-Katz embedding}
\label{KKSec}

In this case we consider the embedding \cite{KarchKatz},
\begin{equation}
\label{Karchf}
f_{KK}=-\frac{(z^1)^2+(z^2)^2}{2}-\mu^2\,,
\end{equation}
which breaks $SO(4)$ down to $SO(2) \times SO(2)$, with the former $SO(2)$ acting on $z^1$ and $z^2$ and the latter acting on $z^3$ and $z^4$.  The supersymmetric vacua consist of two disjoint $S^1$'s, where on each $S^1$ one $SO(2)$ acts naturally while the other is trivial; as with the Kuperstein case some of the symmetry leaves the supersymmetric vacua fixed, but unlike that case here there is a moduli space generated by the rest of symmetry.  Likewise, as with the Kuperstein case we shall see that the nonsupersymmetric vacua nontrivially realize all of the symmetry, and for this embedding the corresponding spaces of vacua will have topology $T^2$.

We have for this embedding
\begin{eqnarray}
\partial_2\zeta=0, \quad \partial_{i}\zeta=-{z^{i}\over n f}\,, \quad i=\{3, 4\}\,,
\end{eqnarray}
and
\begin{eqnarray}
\label{KarchG}
G={R^2(\epsilon^2-R^2)\over \Gamma |f|^2}\,,
\end{eqnarray}
where $R^2 \equiv (z^1)^2 + (z^2)^2$.
For this example we will proceed directly to the long throat and large modulus limit, where we have
\begin{eqnarray}
\label{Karchdelta}
\delta =-{a e^{4u}+2\over a e^{4u}+4} G\approx -G\,.
\end{eqnarray}
The first derivatives can be calculated to be
\begin{eqnarray}
\partial_2 V&=&0 \,, \\
\label{KarchVi}
{\partial_{i} V}&=& {|W_0|^2 a z^i \over \omega^2 3 e^{8u}n f} \left(-{2\over 3}-\delta-{1 \over 3}{f\over \bar{f}}+{n\over \Gamma \bar{f}}(2R^2-\epsilon^2)+{(n-1)R^2(\epsilon^2-R^2)\over \Gamma |f|^2}\right)=0\,.
\end{eqnarray}
We can combine equations (\ref{KarchG}), (\ref{Karchdelta}), and (\ref{KarchVi}), giving the complex equation 
\begin{equation}
\label{KarchComplexEqn}
-{2\over 3}|f|^2-{1 \over 3}f^2+{n\over\Gamma} \left(\mu^2 (2R^2 - \epsilon^2) + {1 \over 2} \epsilon^2 R^2\right)=0\,,
\end{equation}
which is clearly quadratic in $R^2$.  As in the Kuperstien case, restricting to the tip of the geometry we find that the equations are generically overconstrained, and we again treat $\mu$ as a tunable parameter in order to find solutions to equation \ref{KarchComplexEqn}.  Considering the imaginary part of \ref{KarchComplexEqn}, we have 
\begin{eqnarray}
\label{KarchIm}
({\rm Im}[\mu^2])\Big( R^2 ({2 n\over \Gamma} - {1 \over 3} ) - {n \epsilon^2 \over \Gamma}- {2 \over 3} {\rm Re}[ \mu^2]\Big)=0\,,
\end{eqnarray} 
giving again two options for constraining the embedding.

\subsection{Real  7-brane embedding}
If we satisfy equation \ref{KarchIm} by forcing ${\rm Im}[\mu^2]=0$, we can solve equation \ref{KarchComplexEqn} to obtain
\begin{eqnarray}
R^2= - 2\mu^2+{n\over\Gamma}(\epsilon^2+4\mu^2) \pm {1\over\Gamma}\sqrt{n(16\mu^4(n-\Gamma)+8\mu^2\epsilon^2(n-\Gamma)+n\epsilon^4)}\,.
\end{eqnarray}
For the choice of positive sign, there is no value of $\mu$ that gives $R^2 \leq \epsilon^2$.  Choosing the minus sign and assuming $\epsilon^2 \leq \mu^2$, we have in the small-$\epsilon$ limit,
\begin{eqnarray}
\label{KKSoln}
R^2\approx {-\Gamma'\mu^2\epsilon^{4/3}\over 2n}+{\epsilon^2\over2}\,.
\end{eqnarray}
In order to confine the solutions to the tip, the first term must not dominate over the second.  Thus we have solutions in the window $\epsilon^2 \leq \mu^2 \leq \epsilon^{2/3}$. These solutions, as stated, have topology $T^2$ for $0 < R^2 < \epsilon^2$; for choices of $\mu$ much less than the maximum $ \mu^2 \ll \epsilon^{2/3}$,  the second term in (\ref{KKSoln}) dominates and the vacua approach the square torus $(z^1)^2 + (z^2)^2 = (z^3)^2 + (z^4)^2 = \epsilon^2/2$.

Since we have $\delta \approx -G$ in this limit, we again have the volume and potential approaching supersymmetric values (\ref{VVSusy}), (\ref{SusyVol}).  Examining the stability of the vacua, we set
\begin{eqnarray}
\mu^2={\beta n \epsilon^{2/3}\over\Gamma'} \quad |\beta|<1\,,
\end{eqnarray}
and the same steps as in equations (\ref{SKup2d}) through (\ref{SKupM/V}) we can calculate 
\begin{eqnarray}
{m^2\over |V|}=\big\{{a^2 e^{8u}\over 3}, {a^2 e^{8u}\over 3}, {9\over5\beta^2}+{1\over\beta}-{22\over15}, 0, 0, {9\over5\beta^2}-{1\over\beta}-{22\over15},  {a\Gamma' e^{4u} (\beta^2-1)\epsilon^{4/3}\over4 n^2\beta^2}, {a\Gamma' e^{4u} (\beta^2-1)\epsilon^{4/3}\over4 n^2\beta^2}\big\} \,.
\end{eqnarray}
Again a small range of $\beta$ leads to stable solutions, as the last two eigenvalues are the most constraining on $\beta$:
\begin{eqnarray}
{1\over 1+{3n^2\over a\Gamma' e^{4u} \epsilon^{4/3}}} \leq \beta^2 < 1.
\end{eqnarray}
  Most solutions are hence unstable, but stable examples do exist at the very upper bound of the allowed values of $\mu^2$; unlike the Kuperstein case, here stability depends on the interplay between the large volume and the small value of $\epsilon$.

%and the same steps as in equations (\ref{SKup2d}) through (\ref{SKupM/V}) we can calculate 
%\begin{eqnarray}
%{m^2\over |V|}=\big\{{a^2 e^{8u}\over 3}, {a^2 e^{8u}\over 3}, {27\over10\beta^2}-{61\over30}, {9\over5\beta^2}-{1\over\beta}-{22\over15}, {9\over10\beta^2}+{1\over 2\beta}-{11\over15}, 0, -{9\over10\beta^2}-{1\over 2\beta}+{11\over15}, {61\beta^2-81\over 30\beta^2}\big\} \,.
%\end{eqnarray}
%A very small range of $\beta$ leads to stable solutions, as the last eigenvalue is the most constraining: $9 \sqrt{2/167} \leq |\beta| < 1$, with $9 \sqrt{2/167} \approx 0.985$.  Most solutions are hence unstable, but stable examples do exist at the very upper bound of the allowed values of $\mu^2$.

\subsection{Complex 7-brane embedding}

If we allow $\mu^2$ to remain complex, to satisfy (\ref{KarchIm}) we must have the constraint
\begin{eqnarray}
{\rm Re}[\mu^2]={-R^2\over2}+ {n(6R^2-3\epsilon^2)\over 2 \Gamma' \epsilon^{4/3}}\,,
\end{eqnarray}
which when combined with equation \ref{KarchComplexEqn} yields
\begin{eqnarray}
R^2={\epsilon^2\over 2} \pm  {\sqrt{3 n \Gamma(3n+\Gamma)(3n\epsilon^4-4 \,{\rm Im}[\mu^2]^2\Gamma)}\over 6n(3n+\Gamma)}\,.
\end{eqnarray}
We see that to keep $R^2$ real we need
\begin{eqnarray}
{\rm Im}[\mu^2]^2=\beta{3n\epsilon^{8/3} \over4\Gamma'}\, \quad \beta \leq 1,
\end{eqnarray} 
so therefore 
\begin{eqnarray}
R^2={\epsilon^2 \over 2} \pm \sqrt{\Gamma'(1-\beta)\over12n}\epsilon^{8/3}\,.
\end{eqnarray}
The first term dominates, giving vacua again very close to the square torus at $R^2 = \epsilon^2/2$.

Calculating the second derivatives, for convenience we define 
\begin{eqnarray}
z^1=z^2=\sqrt{R^2 \over 2}\, \quad z^3=z^4=\sqrt{\epsilon^2-R^2\over 2}
\end{eqnarray}
this allows us to determine the eigenvalues as 
\begin{eqnarray}
{m^2\over|V|}=\big\{ {a^2 e^{8u}\over3}, {a^2 e^{8u}\over3}, {16n\over 5\Gamma' \epsilon^{4/3}}, {12n\over 5\Gamma' \epsilon^{4/3}}, {12n\over 5\Gamma' \epsilon^{4/3}}, 0, 0, {-4n\over \Gamma' \epsilon^{4/3}} \big\}
\end{eqnarray}
The final eigenvalue is large and negative, and hence these solutions are unstable.

We have seen that the Karch-Katz case is closely analogous to the Kuperstein case, with nonsupersymmetric vacua appearing at certain restricted values of $\mu^2$ and filling out a two-dimensional moduli space, in this case $T^2$; stability of these AdS vacua is possible but not inevitable.

\section{Vacua for antibranes}
\label{AntibraneSec}

Everything we have discussed so far applies to the potential felt by D3-branes in the presence of nonperturbative moduli stabilization.  We now argue that for $\D3$-branes, the solutions at the tip of the warped throat are the same.

If we consider the nonperturbative effects on the $\D3$-branes at the tip of the throat we know that the source for gravity remains unchanged, but the sign of $C_4$ charge is opposite that of D3-branes.  This is equivalent \cite{DMSU} to flipping the sign of the imaginary part of $\zeta$, or $\zeta(Y) \leftrightarrow \bar{\zeta}(\bar{Y})$. In the two examples we considered we had the embedding function of the form
\begin{eqnarray}
f(z^a)=g(z^a)+\mu\,,
\end{eqnarray}
where $g(z^a)$ is polynomial in the $z^a$ with real coefficients, and for the KK case we conventionally write $\mu^2$ instead of $\mu$.  Since at the tip the $z^a$ are real, it follows that $\mu$  is the only complex variable in our entire analysis. If we now consider the $\partial_a V$ equation (\ref{aDerivEqns}) and take the real and imaginary parts, we see that we must have the general forms 
\begin{eqnarray}
{\rm Im}[\partial_a V]= {\rm Im}[\mu] \, H(z,{\rm Re}[\mu], {\rm Im}[\mu]^2)\, \quad {\rm Re}[\partial_a V]=H'(z,{\rm Re}[\mu],{\rm Im}[\mu]^2)\,,
\end{eqnarray}
for some functions $H$ and $H'$.
Therefore if have $\D3$-branes, where ${\rm Im}[\mu]\rightarrow-{\rm Im}[\mu]$, the equations for $\partial_a V =0$ are unchanged.  Implicit here is the assumption that the axion $b$ has adjusted itself to take whatever value necessary to assure $\delta$ is real.   We thus find the same locations of the antibranes as in the brane case.

\section{Conclusions}
\label{ConclusionsSec} 

We have obtained the general equations for nonsupersymmetric D3-brane vacua in  type IIB flux compactifications stabilized by nonperturbative effects, and solved and analyzed them in the particular cases of two specific 7-brane embeddings at the tip of the warped deformed conifold throat.  Nonsupersymmetric vacua exist in general, but to localize them at the tip requires one tuning of the 7-brane embedding; $\D3$-branes were shown to have the same solutions.

In the large volume, long throat limit, these vacua have a few general interesting properties.  Both the overall compact volume and the value of the effective four-dimensional cosmological constant approach the corresponding values in the supersymmetric case in this limit.  This implies that the precise location of the D3-brane has little effect on the overall volume, as seems intuitively reasonable, and that the nonsupersymmetric vacua are anti-de Sitter solutions.

The vacua in general come in continuous families, corresponding to the orbits of the geometric symmetries preserved by the 7-brane embedding.  Interestingly, the nonsupersymmetric vacua we have obtained come in higher-dimensional spaces than the corresponding supersymmetric vacua.  We have also analyzed the question of stability, comparing the eigenvalues of the mass matrix to the Breitenlohner-Freedman bound, and found that (perturbative) stability for nonsupersymmetric vacua is possible though not generic.  Nonetheless, the existence of stable nonsupersymmetric vacua for branes in warped throats is worth remarking upon.  Although the equations are considerably more difficult, the existence of nonsupersymmetric vacua off the tip should be expected and would be interesting to investigate.  Furthermore, having mapped out further the landscape of brane vacua in warped throats, further investigation of the consequences for dynamics such as brane inflation would be of considerable interest.

\bigskip\bigskip\bigskip

\centerline{\bf{Acknowledgements}}

\bigskip
We would like to thank Tom DeGrand, Shanta de Alwis, Liam McAllister and Ben Shlaer for discussions.  O.D. would like to acknowledge the Banff International Research Station where this work was completed.  This work was supported by the DOE under grant DE-FG02-91-ER-40672.

\end{document}